\documentclass{article}
\usepackage{amsmath}
\usepackage[nopatch]{microtype}
\usepackage{booktabs}
\usepackage{kotex}
\usepackage{authblk}

\usepackage[utf8]{inputenc}
\usepackage{geometry}
\usepackage{algorithm}
\usepackage{algpseudocode}
\usepackage{multicol}
\usepackage{varwidth}
\usepackage{graphicx}

\title{Efficiency-Improved Inter-Rollup Transfer System Leveraging \\  Batch Settlement Methods}

\author[1]{Hyun Jeong \thanks{tariz@theradius.xyz}}
\author[2]{Hyemin Lee \thanks{lhm0710@snu.ac.kr}}
\affil[1]{Radius Lab}
\affil[2]{Radius Lab and Seoul National University}
\begin{document}
\maketitle

\begin{abstract}
As the significance of blockchain innovation grows and the focus on scalability intensifies, rollup technology has emerged as a promising approach to tackle these scalability concerns. Nonetheless, rollups encounter restrictions when interacting with other rollups, leading to diminished throughput, increased latency, higher fees, and a complex user experience in transactions between rollups. In this paper, we put forth a novel system that employs batch settlement techniques to augment the efficiency of transfers between rollups. Our proposed system comprises a settlement rollup responsible for batch settling transfers among rollups and a smart contract structure that carries out the settlements. Notably, we utilize a zero-knowledge proof algorithm to guarantee the computational integrity of the settlement rollup while ensuring security through Ethereum smart contracts for proof verification and settlement execution. By implementing this approach, the proposed system can effectively and securely execute asset transfers between rollups, ultimately improving their scalability and usability. Consequently, our research provides a fresh perspective on resolving the challenges of throughput, latency, and fees associated with transfer systems.

\end{abstract}
% \keywords{keyword entry 1, keyword entry 2, keyword entry 3} %% First letter not capped

\begin{multicols}{2}

\section{Introduction}

Blockchain technology has gained considerable attention due to its potential to revolutionize the financial industry by offering transparency and reliability in transactions, as well as facilitating services such as international money transfers and real-time payments\cite{ref1}. As prominent blockchains like Ethereum experience a surge in popularity, an increase in both users and transaction volumes has highlighted the pressing issue of blockchain scalability\cite{ref2}. This scalability challenge has manifested itself in the form of limited transaction throughput, increased latency, and higher transaction fees, ultimately leading to reduced usability and hindering the widespread adoption of blockchain technology\cite{ref3}.

In an effort to tackle these challenges, rollup technology has emerged as a promising solution for enhancing blockchain scalability\cite{ref4}. Rollups work by processing transactions off-chain and only committing the outcomes to the blockchain, effectively increasing its throughput capabilities. The adoption of rollup technology can alleviate the computational strain on the blockchain, accelerate transaction processing times, and decrease transaction fees\cite{ref5}, \cite{ref6}.

Recognizing these advantages, blockchain-based applications are increasingly leveraging rollup technology to build their own networks, aiming to deliver highly-scalable services and improve overall usability\cite{ref7}, \cite{ref8}, \cite{ref9}, \cite{ref10}. However, since rollups operate independently within separate off-chain environments, they face challenges with direct interactions between different rollups\cite{ref11}, \cite{ref12}. For example, transferring assets between rollups on the Ethereum blockchain necessitates the use of the Ethereum blockchain as an intermediary. Consequently, all transfers between rollups must be executed on the Ethereum blockchain. This reliance on individual settlements, limited by the blockchain's throughput, results in longer waiting times and additional fees for rollup users, negatively impacting usability\cite{ref13}. These issues may not only impede the growth of blockchain-based applications but also adversely affect the adoption and expansion of the blockchain ecosystem.

In this paper, we present an innovative approach not previously explored in existing literature or research to tackle the challenges faced in inter-rollup transactions. Our proposed system enhances the efficiency of blockchain transactions by employing batch settlements, which consolidate multiple transactions for processing at once, rather than individually settling them across rollups. The system incorporates a settlement rollup for handling batch settlements and a smart contract structure for executing payments.

Moreover, we introduce a separate procedure to validate the proper operation of the batch settlement within the settlement rollup. To accomplish this, a zero-knowledge proof algorithm is utilized, which is capable of verifying the computational integrity of the settlement rollup[14]. Additionally, the Ethereum blockchain-based smart contracts are employed to provide transparency and security for settlement payments by performing validity verification of batch settlements. By proposing a novel inter-rollup transfer system that considers the scalability of blockchain, we anticipate an improvement in the usability of rollups.

The structure of this paper is as follows: Chapter 2 introduces the fundamental concepts of blockchain technology and rollups. Chapter 3 analyzes the issues with individual settlements in inter-rollup transfers, while Chapter 4 proposes a new protocol utilizing batch settlements. Chapter 5 presents the implementation of the proposed protocol, and finally, Chapter 6 offers conclusions and future research directions.

\section{Background}

\subsection{Blockchain}

Blockchain technology is a form of distributed ledger technology that provides a system for multiple participants in a peer-to-peer network to securely store and share data without the need for a central administrator or centralized server\cite{ref15}. A blockchain is composed of a series of interconnected blocks, with each block containing the hash value of the previous block, creating a linkage. This structure contributes to ensuring the data integrity within the blockchain\cite{ref16}.

With the evolution of blockchain technology and the widespread use of prominent blockchains like Ethereum, there has been a significant increase in transaction volume. This surge has emphasized the scalability challenges faced by blockchains, as the capacity for handling transactions at the base layer is limited. In the context of blockchain, scalability refers to a system's ability to efficiently manage high levels of usage and transaction demand\cite{ref2}. Insufficient blockchain scalability can lead to slower transaction processing speeds and increased fees, ultimately impeding the growth of blockchain-based applications and the technology's mass adoption\cite{ref3}.

To tackle these scalability issues, a variety of solutions are being researched and developed. These include rollups, state channels, plasma, and sharding, all of which are proposed solutions aimed at increasing transaction throughput, enhancing processing speeds, and overcoming the constraints inherent in traditional blockchain systems\cite{ref17}.

\subsection{Smart contract}

Smart contracts, as automated digital agreements built on blockchain technology, are comprised of program code that automatically executes when specific contractual conditions are fulfilled\cite{ref18}. These smart contracts obtain computational integrity from the consensus algorithm and distributed ledger structure of the blockchain network, thereby ensuring the dependability of their results. Notably, each participant in the blockchain network maintains an identical copy of the smart contract, with any changes being processed consistently in accordance with the consensus algorithm. This approach secures the trustworthiness of the smart contract's computations and deters potential exploitation by malicious actors\cite{ref19}, \cite{ref20}.

Additionally, smart contracts can contribute to the security and transparency of scalability solutions\cite{ref21}, \cite{ref22}. For instance, Ethereum blockchain smart contracts can validate the proofs submitted with batch transactions in rollups, thereby safeguarding the integrity of the rollup's computational outcomes.

In this paper, smart contracts are leveraged to ensure computational integrity in the settlement rollup and to deliver security and transparency for settlement payments. Specifically, the Ethereum smart contract is responsible for verifying the validity of the batch settlement in the settlement rollup and processing the settlement balance only when the results are deemed valid.

\subsection{Rollup}

Rollups have emerged as a solution to tackle the scalability issues in blockchains, overcoming the constraints of traditional blockchain processing capacity by aggregating transactions executed off-chain and subsequently recording their results on-chain. Rollup operators are responsible for gathering transactions from users, processing these transactions, and submitting the outcomes to the blockchain. The Ethereum blockchain, through the use of smart contracts, validates the submitted results and records them if found to be accurate. During this entire process, rollup operators ensure data availability for off-chain transactions and guarantee proper transaction processing in line with the rollup protocol\cite{ref6}.

The method used for result verification depends on the specific rollup technology employed. Optimistic rollups store compressed transaction results on the blockchain, allowing network participants to confirm the validity of the transaction processing outcomes\cite{ref23}. ZK-rollups, on the other hand, utilize zero-knowledge proof techniques to create compressed proofs for transaction processing results, which are then verified and stored on the blockchain\cite{ref24}. These verification approaches guarantee the integrity and legitimacy of off-chain transactions within rollups\cite{ref14}.

Rollups employ smart contracts to facilitate interaction with the Ethereum blockchain, which are used for depositing or withdrawing funds and ensuring the security and transparency of rollups. Specifically, smart contracts in rollups serve the following purposes\cite{ref11}, \cite{ref22}:

\begin{enumerate}
    \item  State management: Rollup smart contracts are responsible for tracking and managing the rollup system's state, which includes information such as account balances. This state information is synchronized with the underlying blockchain data.
    \item  Deposit handling: To transfer assets from the blockchain to the rollup, users deposit assets into the rollup smart contract. Upon verifying the deposit, the smart contract updates the rollup state through the rollup operator.
    \item  Withdrawal handling: To move assets from the rollup back to the blockchain, users initiate withdrawals through the rollup smart contract. The smart contract verifies the withdrawal request and updates the rollup state via the rollup operator.
    \item  Batch verification: The smart contract is responsible for verifying proofs submitted with batch-processed transactions in the rollup, and updating the rollup state based on this information.
    \item  Security and dispute resolution: The smart contract enforces mechanisms that maintain the security and reliability of the rollup system, ensuring its overall safety.
\end{enumerate}

Rollups contribute to the enhancement of blockchain processing capacity while maintaining its security features. Recent studies have shown that advancements in rollup technology play a significant role in addressing blockchain scalability challenges\cite{ref25}.

In this paper, we present a rollup architecture that capitalizes on off-chain batch settlements for transactions between different rollups. By leveraging smart contract interactions, our approach seeks to improve the efficiency of transfers across rollups.

\subsection{Zero-knowledge proof}

Zero-knowledge proofs refer to a technique that enables a prover to establish the truthfulness of a statement without divulging any information. Typically, a zero-knowledge proof must meet the following three criteria\cite{ref26}:

\begin{enumerate}
    \item  Completeness: If a valid proof is supplied, the verifier will accept the prover's assertion.
    \item Zero-knowledge: Although the verifier acknowledges the truth of the prover's statement, they obtain no additional information in the process.
    \item Soundness: In the event that the prover presents a false statement, the likelihood of the verifier accepting the claim is extremely low.
\end{enumerate}

Within a zero-knowledge proof system, the proof generation phase entails the prover convincing the verifier of the accurate execution of a specific computation without exposing any confidential information. Throughout this stage, the prover constructs a legitimate proof for a designated circuit, while the verifier assesses this proof to ascertain the precision of the computation. The primary steps involved are as follows\cite{ref27}:

\begin{enumerate}
    \item Circuit definition: Develop a circuit that embodies the problem to be proven. This circuit is capable of executing intricate calculations, encompassing logical, arithmetic, and bitwise operations.
    \item Proof generation: Utilizing the circuit's inputs and private data, the prover formulates a zero-knowledge proof. The prover refrains from directly revealing their confidential information, instead supplying only the essential public details for proof generation.
    \item Proof verification: The verifier examines the zero-knowledge proof acquired from the prover, confirming the proper execution of the computation. During this verification phase, the verifier can authenticate the validity of the proof without directly accessing the prover's private data.
\end{enumerate}

Zero-knowledge proofs play a crucial role in maintaining the integrity of off-chain processing results, thereby addressing the scalability challenges faced by blockchains. In this process, a prover generates a mathematical proof for off-chain transactions. This proof serves to confirm the correctness of transaction processing without divulging any details about the transactions themselves. Subsequently, smart contracts act as verifiers, employing zero-knowledge proof algorithms to authenticate the provided proof. Upon successful validation, the legitimacy of off-chain transaction processing can be assured for all relevant parties\cite{ref28}, \cite{ref29}.

In this paper, we leverage zero-knowledge proofs to enable smart contracts to produce evidence that attests to the validity of batch settlements within settlement rollups, thus guaranteeing the computational integrity of these settlements.

\subsection{Interbank transfer}

One method of providing a novel perspective on the transfer process between rollups is to analyze the contemporary interbank transfer system. When Alice, a customer of Bank A, intends to send funds to Bob, a customer of Bank B, Bank A first reviews and approves Alice's transfer request. Then, Bank A sends the transfer request to Bank B using an interbank transfer system, such as SWIFT. Upon receiving the request, Bank B transfers the funds to Bob's account. Throughout this procedure, there is no direct transfer of funds between Bank A and Bank B; instead, settlements occur through an intermediary, like a central bank\cite{ref30}.

Both Bank A and Bank B provide transfer details to an intermediary institution, such as a central bank, to request settlements. This information includes the transfer amount, fees, and transfer date. The intermediary verifies the details, consolidates the comprehensive transaction history between the two banks, and calculates each bank's settlement balance. Settlement funds are then transferred between the banks using the intermediary's transaction accounts. Upon settlement completion, the intermediary supplies both Bank A and Bank B with the final transaction information and settlement results\cite{ref31}.

In this paper, we put forth an efficient inter-rollup transfer system, drawing inspiration from existing interbank transfer and settlement techniques.

\section{Motivation}

Rollups are networks functioning autonomously within separate off-chain environments. Each rollup employs distinct computing and storage infrastructure to process transactions, resulting in data isolation, incompatible authentication and consensus mechanisms, and increased communication overheads. These factors make direct interactions between rollups difficult, necessitating the use of the Ethereum blockchain for asset movement[11], [12].

\begin{figure}[H]
\centering
\includegraphics[width=1\linewidth]{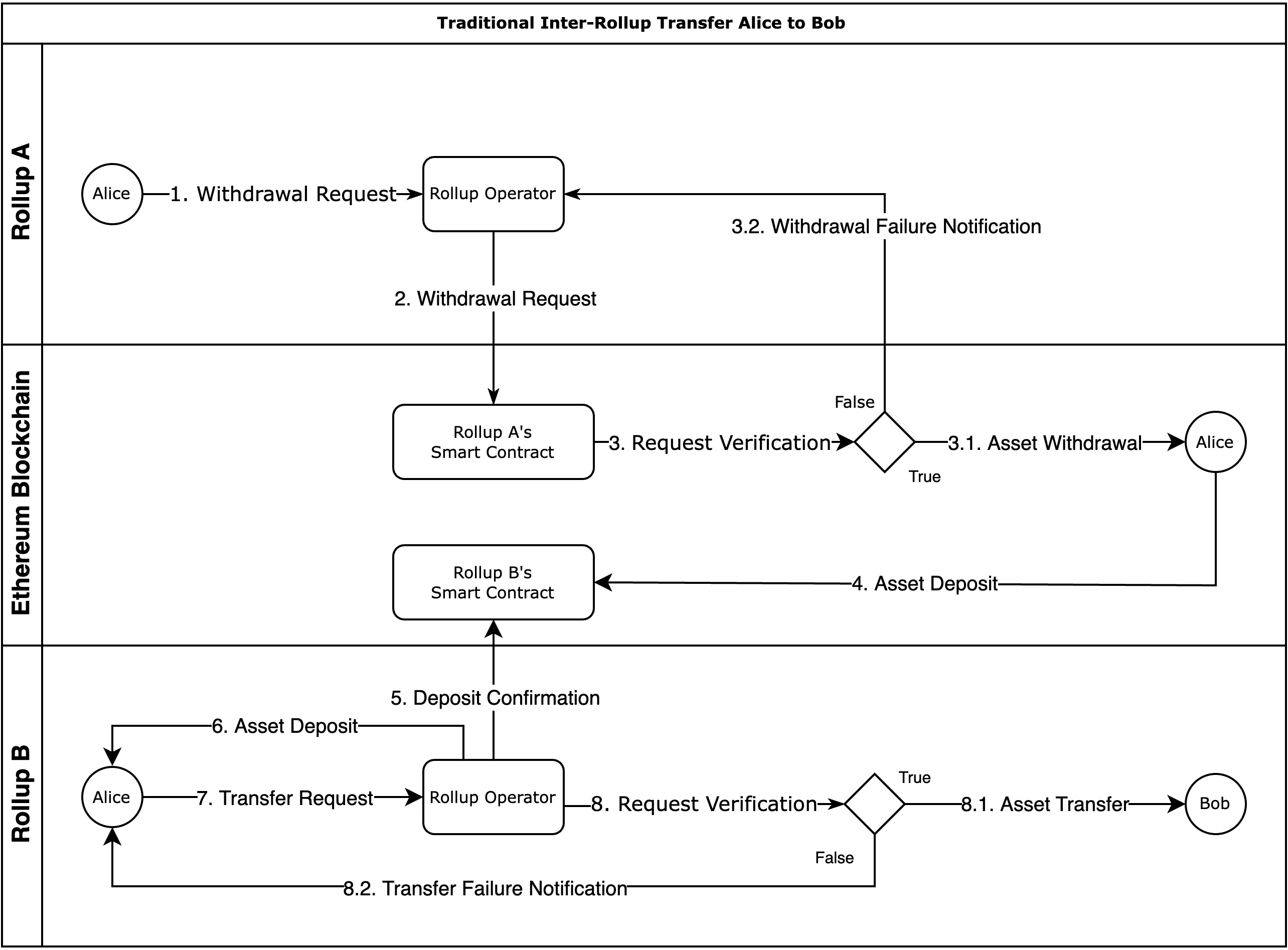}
\caption{Traditional Inter-Rollup Transfer Workflow}
\label{fig_sim}
\end{figure}

Consider a scenario where Alice, a customer of Rollup A, wants to transfer funds to Bob, a customer of Rollup B, via the Ethereum blockchain\cite{ref32}, \cite{ref33}. Figure 1 demonstrates the process unfolds as follows:

\begin{enumerate}
    \item  Alice requests the Rollup A operator to withdraw assets.
    \item  The Rollup A operator submits a withdrawal request to Rollup A's smart contract on the Ethereum blockchain.
    \item  Rollup A's smart contract verifies Alice's withdrawal request.
        \begin{enumerate}
            \item True: Assets are sent to Alice's Ethereum blockchain address. 
            \item False: The operator receives notification of the failed withdrawal.
        \end{enumerate}
    \item  Alice deposits assets into rollup B's smart contract on the Ethereum blockchain.
    \item  The Rollup B operator verifies Alice's asset deposit in Rollup B's smart contract.
    \item  The Rollup B operator deposits assets to Alice's address in Rollup B.
    \item  Alice requests the Rollup B operator to transfer assets to Bob's account.
    \item  The Rollup B operator validates the request.
        \begin{enumerate}
            \item True: Assets are transferred to Bob's address.
            \item False: Alice is notified of the unsuccessful withdrawal. 
        \end{enumerate}
\end{enumerate}

As previously discussed, for asset transfers between rollups, it is crucial to perform direct asset movements through each rollup's smart contracts on the blockchain. This requirement stems from the need to verify the legitimacy of asset transfers via settlements on the blockchain. Since each rollup employs distinct data authentication techniques, trusting each other's information and calculations can be challenging. Nevertheless, because the data and processing outcomes stored on the mutually linked blockchain are reliable, individual settlement procedures are conducted during asset transfers. Although the blockchain guarantees the security of these processes, a minimum of three blockchain transactions for a single transfer between rollups imposes limitations on the processing capacity of the blockchain for rollup usage\cite{ref5}, \cite{ref6}.

Rollups are a technology that blockchain-based applications utilize to enhance usability and scalability. However, the restricted interactions between rollups result in reduced usability due to factors such as transaction speed constraints, elevated fees, and the necessity for complicated usage processes\cite{ref3}. Specifically, this structure hampers transaction liquidity by making it difficult to use multiple rollups concurrently. These issues may create obstacles for the growth of blockchain-based applications and negatively impact blockchain adoption and ecosystem expansion\cite{ref13}.

This paper presents a novel system designed to resolve these challenges by improving inter-rollup interactions. The proposed system incorporates batch settlement from traditional banking systems to boost the efficiency of transfers between rollups. In particular, the participating rollups collaboratively carry out the verification of batch settlements and process the settlement balances within a trusted Ethereum blockchain smart contract. This approach ensures both efficiency and security in inter-rollup transfers.

\section{Design}

\subsection{Overview}

\begin{figure}[H]
\centering
\includegraphics[width=1\linewidth]{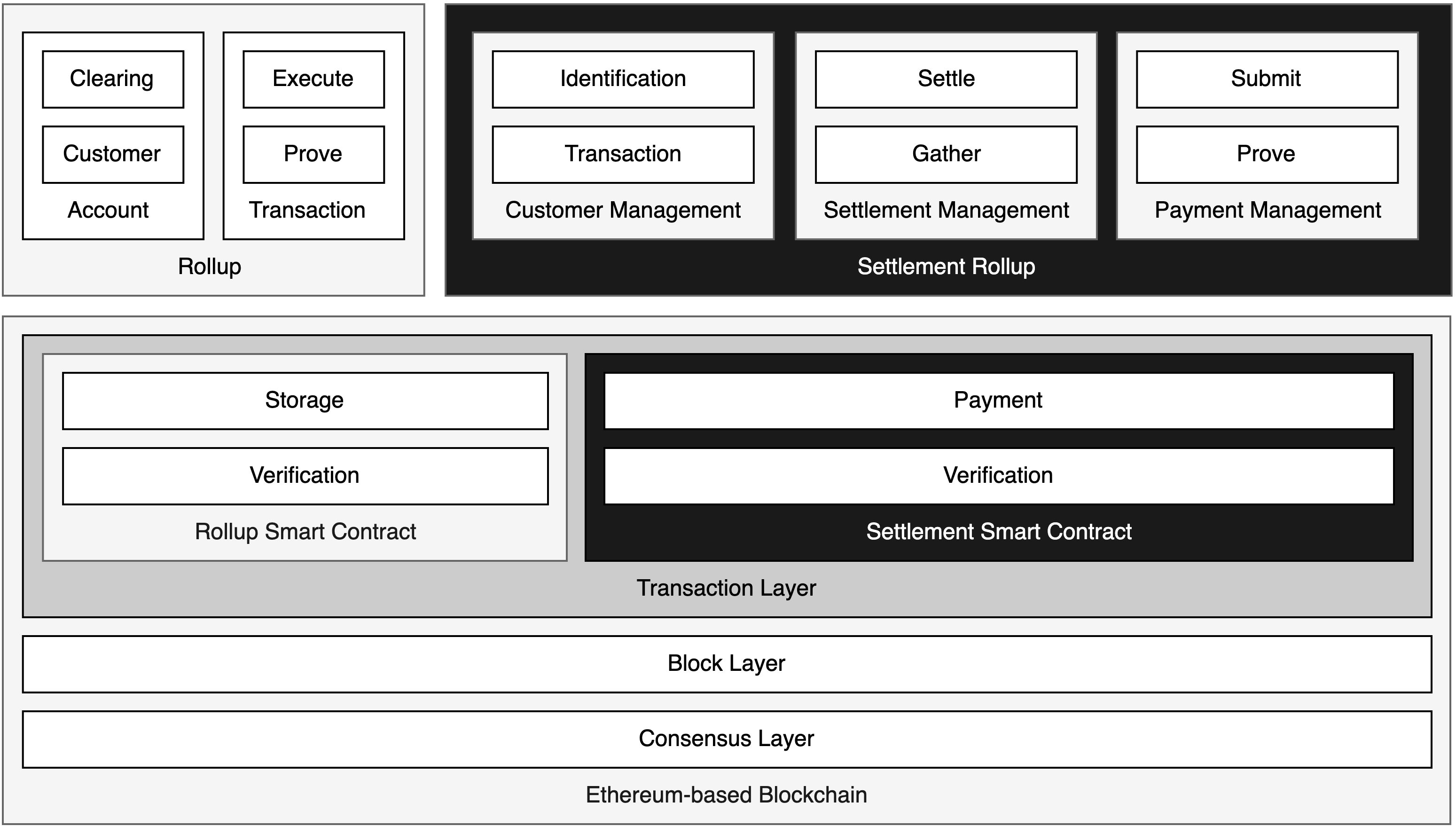}
\caption{Comprehensive Architecture of the Inter-Rollup Transfer System}
\label{fig_sim}
\end{figure}

The proposed inter-rollup transfer system's overall structure is depicted in Figure 2. This system comprises three layers: the rollup, the settlement rollup, and the settlement smart contract. Rollups are off-chain components connected to the blockchain, periodically submitting transaction execution results to their corresponding smart contracts on the blockchain. The settlement rollup serves as an intermediary, processing transfer records between rollups and forwarding the outcomes to the settlement smart contract. The settlement smart contract validates the settlement results received from the settlement rollup and records the data on the blockchain. Moreover, the settlement smart contract manages transaction accounts for the transfer of settlement funds among rollups.

\subsection{Settlement rollup}

The settlement rollup identifies rollup information, collects and settles rollup transfer data, interfaces with the smart contract to request settlement balance payments, and ultimately delivers the settlement results to the rollup.

\subsubsection{Customer management}
    
Customer management involves identifying rollups utilizing the inter-rollup transfer system and managing the requisite information for settlement. The settlement rollup oversees three types of information for customers using the inter-rollup transfer system: rollup ID, settlement account, and clearing account.

The rollup ID is used to identify customers employing the inter-rollup transfer system. Each rollup possesses a unique chain ID on the blockchain. The settlement rollup sets this as the rollup ID and stores it in a database. This information is employed to confirm the requesting rollup's customer status during settlement requests.

The settlement account is designated for rollups to store, send, and receive settlement funds. As settlements are processed in the settlement smart contract, settlement accounts are created as Ethereum blockchain addresses. Upon requesting customer registration from the settlement rollup, the settlement rollup interacts with the settlement smart contract to generate the rollup's settlement account and storages this information in a database. The rollup is then provided with the settlement account address.

The clearing account is employed by rollup users to transfer or receive funds from users in other rollups. This account is an internal rollup account that does not interact with the blockchain. Rollups establish clearing accounts for other connected rollups through the inter-rollup transfer system. For instance, if Rollup A and Rollup B utilize the inter-rollup transfer system, both rollups create clearing accounts for each other. Thus, Rollup A creates a clearing account for Rollup B, and Rollup B creates a clearing account for Rollup A. Likewise, if Rollup A and Rollup C use the inter-rollup transfer system, Rollup A establishes a clearing account for rollup C. Consequently, Rollup A maintains two clearing accounts (B and C).

Upon creating the clearing accounts, the rollup notifies the settlement rollup of the account information. The settlement rollup storages this data in a database, which is subsequently used to calculate balances between rollups during settlement requests.

\subsubsection{Settlement management}
    
Settlement management is the process of gathering and organizing transaction information between rollups to compute settlement balances. During settlement, the settlement rollup acquires all required transaction details from the involved rollups. This information includes the rollup ID, sender, receiver, clearing account, transaction amount, and transaction timestamp.

\begin{figure}[H]
\centering
\includegraphics[width=1\linewidth]{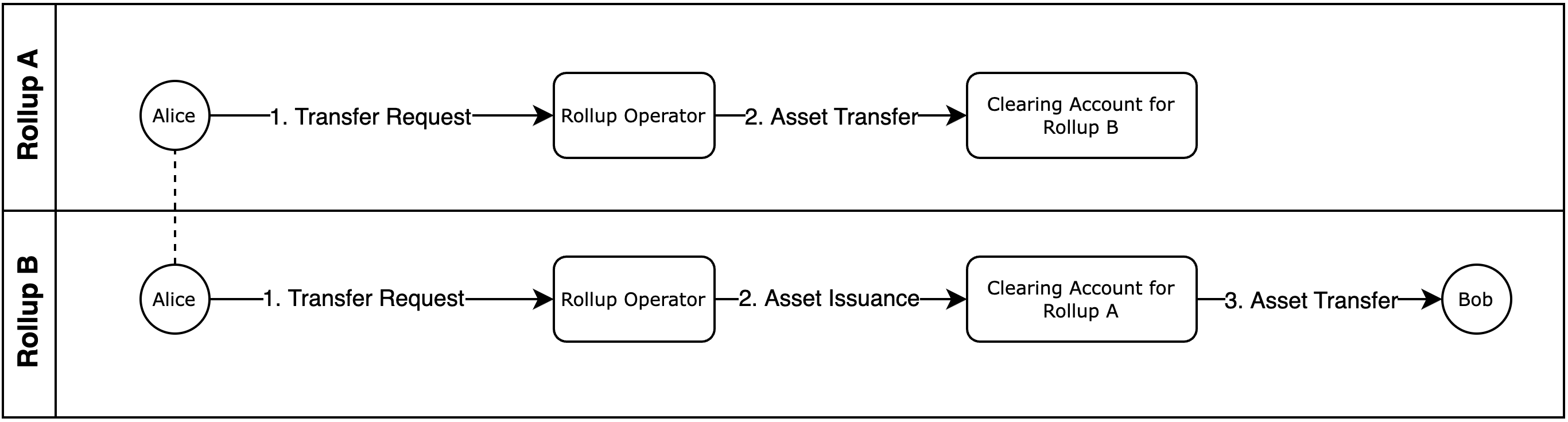}
\caption{Modifications in Clearing Account States}
\label{fig_sim}
\end{figure}    

The settlement rollup matches and reconciles transactions between the sender and receiver rollups to ensure their accuracy and completeness. The provided Figure 3 demonstrates the changes in clearing accounts during inter-rollup transfers. For instance, consider Alice from Rollup A sending 10 assets to Bob in Rollup B. Initially, Alice transfers 10 assets to Rollup A's clearing account for Rollup B. Then, Rollup B issues 10 assets from its clearing account for Rollup A and sends them to Bob. Consequently, the balance in Rollup A's clearing account for Rollup B increases by 10 assets, while the balance in Rollup B's clearing account for Rollup A decreases by 10 assets. The settlement rollup verifies the transaction's accuracy by ensuring that the sum of the matched clearing account balances equals zero.

The settlement rollup calculates each rollup's settlement balance and generates a transaction that allows the settlement smart contract to process the settlement funds. For example, suppose Rollup A's clearing account for Rollup B has a balance of 10 assets, and Rollup B's clearing account for Rollup A has a balance of 5 assets. The settlement balance is computed as the difference between the clearing account balances. Therefore, since Rollup A's settlement balance is 5 assets and Rollup B's settlement balance for Rollup A is 0 assets, Rollup A must transfer 5 assets to Rollup B in the final transaction.

Algorithm 1 presents the settlement rollup's process of reconciling transactions and computing settlement balances. Initially, the settlement rollup is assumed to have collected transaction lists from each rollup. Step 1 filters transactions containing clearing account addresses to id-

\begin{algorithm}[H]
\scriptsize
\caption{Execution of settlements by the settlement rollup}\label{alg:cap}
\begin{algorithmic}
\Ensure{each rollup sent transaction data to settlement rollup}
\Procedure{GenerateValidityProof}{rollupTxsList}
\State clearingAccountList $\gets$ rollups' clearing accounts
\State settleTxsList $\gets$ filtered transactions to settle
\newline
\Comment{Step 1: filter transaction to settle} \\
\begin{varwidth}{\linewidth}
\ForAll{rollupTxs $\in$ rollupTxsList}
    \ForAll{rollupTx $\in$ rollupTxs}
        \State sender $\gets$ sender addresss from rollupTx
        \State recipient $\gets$ recipient address from rollupTx
            \If{sender $\in$ clearingAccountList $\lor$ \par\hskip\algorithmicindent \hskip\algorithmicindent \hskip\algorithmicindent \hskip\algorithmicindent 
            recipient $\in$ clearingAccountList}
                \State settleTxsList $\gets$ settleTxsList $\cup$ rollupTx
            \EndIf
    \EndFor
\EndFor
\end{varwidth}
\newline
\Comment{Step 2: settle and generate validity proof} \\
\begin{varwidth}[t]{\linewidth}
\State proof, result $\gets$ \Call{proveSettlement}{settleTxsList}
\end{varwidth}
\newline
\Comment{Step 3: generate tx for layer 1}
\State verifyTx $\gets$ \Call{generateTx}{proof, result}
\State \textbf{return} verifyTx
\EndProcedure
\\
\Procedure{proveSettlement}{filteredTxsList}
\State circuit $\gets$ A zk circuit for settlement 
\State proof, settlementResult $\gets$ \Call{prove}{circuit, settleTxsList}
\State \textbf{return} proof, settlementResult
\EndProcedure
\end{algorithmic}
\end{algorithm}

$ $\\entify those eligible for settlement. In Step 2, the proveSettlement() function reconciles transactions within the settlement circuit, calculates the total settlement amount, and generates proof for it. More details are provided in Algorithm 2. In Step 3, if the proof is valid, the generateTx() function creates a transaction to be executed by the settlement smart contract. This transaction includes requests for transferring and receiving settlement balances from the participating rollups' settlement accounts.

\subsubsection{Payment management}

Payment management involves the settlement rollup generating a proof of the settlement results' validity and submitting it, along with the settlement results, to the smart contract. The settlement rollup employs zero-knowledge proof algorithms to prove the validity of the off-chain settlement results. Generally, to validate computation results, all related transactions must be re-executed to confirm identical outcomes. However, using zero-knowledge proofs, smart contracts can determine whether the settlement rollup meets specific conditions without directly computing all transaction state changes.

Algorithm 2 outlines the process of generating a proof to verify settlement results. All steps below are executed within the circuit. Step 1 compares the entire transaction
    
\begin{algorithm}[H]
\scriptsize
\caption{Generation of zero-knowledge proofs for settlements}\label{alg:cap}
\begin{algorithmic}
\Procedure{SettlementCircuit}{settleTxsList}
\newline
\Comment{Step 1: check each transaction's address and amount}
\State totalSent $\gets$ total amount sent from one rollup to another
\State clearingBalances $\gets$ each clearing account's balance
\begin{varwidth}[t]{\linewidth}
\State settlementAmounts $\gets$ total amount to be settled from \par\hskip\algorithmicindent \hskip\algorithmicindent  one rollup to another
\end{varwidth}
\ForAll {settleTx $\in$ settleTxsList}
    \State pairTx $\gets$ find pair tx from settleTxsList
    \State add equal gate(settleTx.recipient, pairTx.recipient)
    \State add equal gate(settleTx.amount, pairTx.amount)
    \State totalSent $\gets$ update total amount
    \State remove pairTx from settleTxsList
\EndFor
\newline
\Comment{Step 2: check clearing account's balance and total transferred amount } \\
\begin{varwidth}[t]{\linewidth}
\For {i $\leftarrow$ 0 ... len(clearingBalance)}
    \For {j $\leftarrow$ 0 ... len(clearingBalance[i])}
    \State add equal gate(clearingBalance[i][j], \par\hskip\algorithmicindent \hskip\algorithmicindent \hskip\algorithmicindent \hskip\algorithmicindent  totalSent[i][j])
\EndFor
\EndFor
\end{varwidth}
\newline
\Comment{Step 3: calculate the amount that rollup($i$) must transfer to another rollup($j$) for settlement } \\
\begin{varwidth}[t]{\linewidth}
\For {i $\leftarrow$ 0 ... len(settlementAmounts)}
    \For {j $\leftarrow$ 0 ... len(settlementAmounts[i])}
     \State {settlementAmounts[i][j] $\gets$  max(0, \par\hskip\algorithmicindent \hskip\algorithmicindent \hskip\algorithmicindent \hskip\algorithmicindent  clearingBalances[i][j] - clearingBalances[j][i])}
     \State add public input(settlementAmounts[i][j])
\EndFor
\EndFor
\end{varwidth}
\EndProcedure
\end{algorithmic}
\end{algorithm}

$ $\\information, ensuring that the senders and receivers in the corresponding transactions match. Step 2 reconciles the transaction amounts of the corresponding transactions to confirm that participating rollups have consistently processed transactions. For example, it verifies that Rollup A transferred 10 assets to the clearing account and Rollup B did not issue 20 assets to Bob instead. Step 3 computes the settlement balances of both rollups, as detailed in Section 4.2.2.

In the settlement rollup process, three key elements are submitted to the settlement smart contract for verification and execution: 1) the transaction list, 2) the validity proof, and 3) the settlement transaction. 

\subsection{Settlement smart contract in ethereum-based \\ blockchain}

Algorithm 3 delineates the procedure for the settlement smart contract to validate the transaction list, assess the proof provided by the settlement rollup, and ultimately settle the balances between rollups.

During Step 1, the verifyRollupTx() function is utilized to ascertain that the transactions used to calculate

\begin{algorithm}[H]
\tiny
\caption{Verification and payment through the settlement smart contract}\label{alg:cap}
\begin{algorithmic}

\Procedure{Settle}{transactionDataList, proof, transferDataList}
\State verification $\gets$ overall verification result
\ForAll {transactionData $\in$ transactionDataList}
    \State byte $\gets$ concatenated byte from transactionData
    \State address $\gets$ rollup's address from transactionData
    \State res $\gets$ \Call{VerifyRollupTx}{address, byte}
    \If{res is False}
        \State verification $\gets$ False
        \State \textbf{break}
    \EndIf
\EndFor
\State verification $\gets$ \Call{VerifyProof}{proof, transferDataList}
\If{verification is True}
   \State \Call{Transfer}{transferDataList}
   \State \Call{ConfirmSettle}{success}
\Else
    \State \Call{ConfirmSettle}{fail}
\EndIf
\EndProcedure
\\
\Procedure{VerifyRollupTx}{address, byte}
\newline
\Comment{Step 1: rollups' transaction data verification}
\State hash $\gets$ \Call{sha256}{byte}
\State finalizedHash $\gets$ call rollup's contract by address
\State \textbf{return} hash == finalizedHash
\EndProcedure
\\
\Procedure{VerifyProof}{proof, publicInputList}
\newline
\Comment{Step 2: validity proof verification}
\State result $\gets$ perform verification operation
\State \textbf{return} result
\EndProcedure
\\
\Procedure{Transfer}{transferDataList}
\newline
\Comment{Step 3: transfer for settlement}
\ForAll {transferData $\in$ transferDataList}
    \State sender $\gets$ sender addresss from transferData
    \State recipient $\gets$ recipient address from transferData
    \State amount $\gets$ transfer amount from transferData
    \State \Call{transferFrom}{sender,recipient, amount}
\EndFor
\EndProcedure
\\
\Procedure{ConfirmSettle}{settlementStatus}
\newline
\Comment{Step 4: event generation}
\If{settlementStatus is Success}
    \State update contract state with recent settlement
    \State \Call{emit}{success}
\Else
    \State freeze contract and execute recovery mechanism
    \State \Call{emit}{fail}
\EndIf
\EndProcedure

\end{algorithmic}
\end{algorithm}

$ $\\the settlement balance are indeed the ones that were executed. Initially, the sha256() hash function is employed to hash the transaction list submitted by the settlement rollup. Subsequently, the stored values from the rollup's smart contract are invoked using the callRollupContract() function. The two hash values are then compared to ensure their identical nature. By leveraging the avalanche effect of the hash function, this verification step confirms the accuracy of the transaction list used in calculating the settlement results.

Step 2 focuses on validating the accuracy of the settlement results computed by the settlement rollup. To achieve this, the verifyProof() function is employed to determine the correctness of the settlement rollup's results. If the verification outcome is positive, the settlement results are deemed valid. Upon successful validation in Step 2, the settlement rollup's transaction is executed in Step 3 via the transfer() function. This allows the settlement smart contract to transfer funds between rollup settlement accounts. Lastly, Step 4 generates a confirmation event through the confirmSettle() function, which serves to inform the rollup that the settlement process is complete.

All statements within Algorithm 3 are executed by nodes participating in the blockchain network, as they are embedded within the smart contract. As a result, nodes reach a consensus each time a statement is executed, either modifying or verifying the state. This approach ensures the integrity of both verification logic and business logic.

\section{Implementation}
This section has detailed the methods employed for managing inter-rollup transfers and settlements using rollups, settlement rollups, and settlement smart contracts.

Table 1 outlines the notations used within the inter-rollup transfer system. \(chainId_A\) and \(chainId_B\) denote rollup IDs that identify rollups utilizing the inter-rollup transfer system, stored in the database. \(account_A\) and \(account_B\) represent the rollups' settlement account addresses, which facilitate the transfer and receipt of settlement funds. \(clearing_{AB}\) refers to the address of the clearing account where funds are deposited when transferred from Rollup A to Rollup B, while \(clearing_{BA}\) is the address of the clearing account for funds transferred from Rollup B to Rollup A. \(address_A\) and \(address_B\) signify the user addresses of rollups engaging with the inter-rollup transfer system. A \(batch\) is a bundle comprising transactions generated by users, including fund transfers and issuances. \(batch_{sig}\) represents a bundle signed by the user, with the signature employing the user's signing key, \(key\). The \(amount\) pertains to the transfer value within inter-rollup transfers. \(settleTx\) and \(settleTx_{sig}\) indicate the transaction and signed transaction, respectively, created by the settlement rollup during the settlement phase. The signature is produced using the settlement rollup's signing key, \(key_{settle}\). Components of the settlement transaction include \(finalizedTxData\), the collected transaction information from rollups; \(filterTxData\), the transaction information filtered based on clearing accounts; \(settleAmount\), the inter-rollup transfer settlement balance; and \(settleProof\), the evidence supporting the validity of the settlement results.

\begin{table}[H]
% \begin{threeparttable}
\caption{Notations for the Inter-Rollup Transfer System}
\label{table_example}
\begin{tabular}{lp{5cm}}
\toprule
Notion & Description\\
\midrule 
\(chainId_{A}\)&Chain ID of Rollup A\\ 
\midrule
\(chainId_{B}\)&Chain ID of Rollup B\\ 
\midrule
\(account_A\)&Settlement Account Address for Rollup A\\ 
\midrule 
\(account_B\)&Settlement Account Address for Rollup B\\ 
\midrule
\(clearing_{AB}\)&Address of the Clearing Account for Funds Transferred from Rollup A to Rollup B\\ 
\midrule 
\(clearing_{BA}\)&Address of the Clearing Account for Funds Transferred from Rollup B to Rollup A\\ 
\midrule
\(address_A\)&User Address in Rollup A\\ 
\midrule
\(address_B\)&User Address in Rollup B\\ 
\midrule
\(batch\)&Bundle Containing User Transactions in Rollup\\
\midrule
\(batch_{sig}\)&Signed Bundle by Rollup User\\
\midrule
\(key\)&Signature Key of Rollup User\\
\midrule
\(amount\)&User's Transfer Amount\\
\midrule
\(settleTx\)&Transaction Containing Settlement Results\\
\midrule
\(settleTx_{sig}\)&Signed Settlement Transaction by Settlement Rollup\\
\midrule
\(key_{settle}\)&Signature Key of Settlement Rollup\\
\midrule
\(fianlizedTxData\)&Finalized Transaction Data Collected by Settlement Rollup from Rollups\\
\midrule
\(filterTxData\)&Filtered Transaction Data Based on Clearing Accounts\\
\midrule
\(settleAmount\)&Settlement Balance of Inter-Rollup Transfers\\
\midrule
\(settleProof\)&Proof for Validating Settlement Results\\
\bottomrule
\end{tabular}
% \end{threeparttable}
\end{table}

\subsection{New rollup registration}

\begin{figure}[H]
\centering
\includegraphics[width=0.9\linewidth]{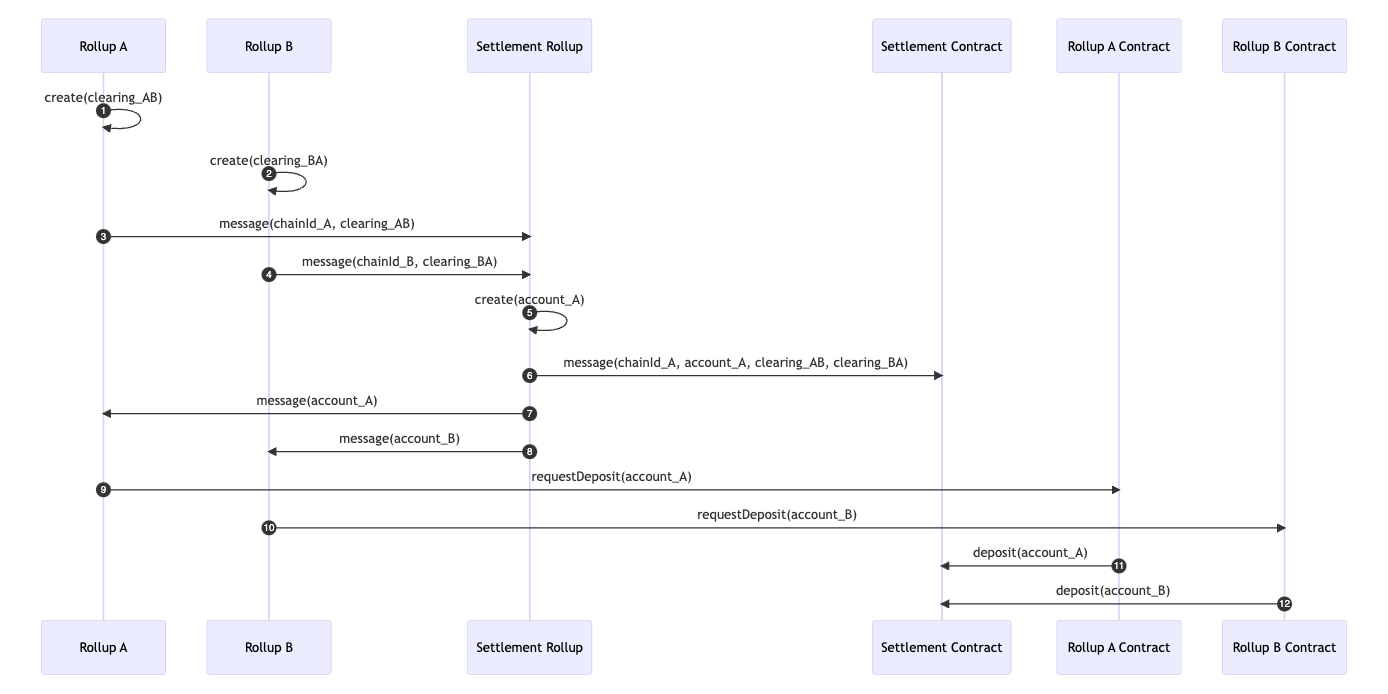}
\caption{Process of Rollup Registration}
\label{fig_wide}
\end{figure}

Figure 4 demonstrates the registration process for a new member, specifically a rollup participating in the inter-rollup transfer system. The settlement rollup retains therollup ID and clearing account details, then generates a settlement account for the rollup. This account is stored and communicated back to the rollup.

To engage in the inter-rollup transfer system, a rollup must submit its chain ID and clearing account information to the settlement rollup. Initially, each rollup creates a clearing account (\(clearing_{AB}\), \(clearing_{BA}\)) designated for the connected rollups within the inter-rollup transfer system. The clearing account information, along with the chain ID (\(chainId_A\), \(chainId_B\)), is then sent to the settlement rollup.

Upon receiving this information, the settlement rollup storages it in a database and creates the rollup's settlement account (\(account_A\), \(account_B\)). The settlement rollup subsequently stores the rollup's chain ID, clearing account address, and settlement account address within the settlement smart contract's storage. The registration result, accompanied by the settlement account details, is then transmitted to the rollup.

Once registration is complete, the rollup requests a deposit into the settlement account via the requestDeposit() function in its smart contract. The rollup's smart contract proceeds to deposit the requested funds into the settlement account, which is accessible by the settlement smart contract using the deposit() function. This deposit is employed to transfer funds to other rollups when handling settlement amounts.

\subsection{Inter-rollup transfer}

\begin{figure}[H]
\centering
\includegraphics[width=0.9\linewidth]{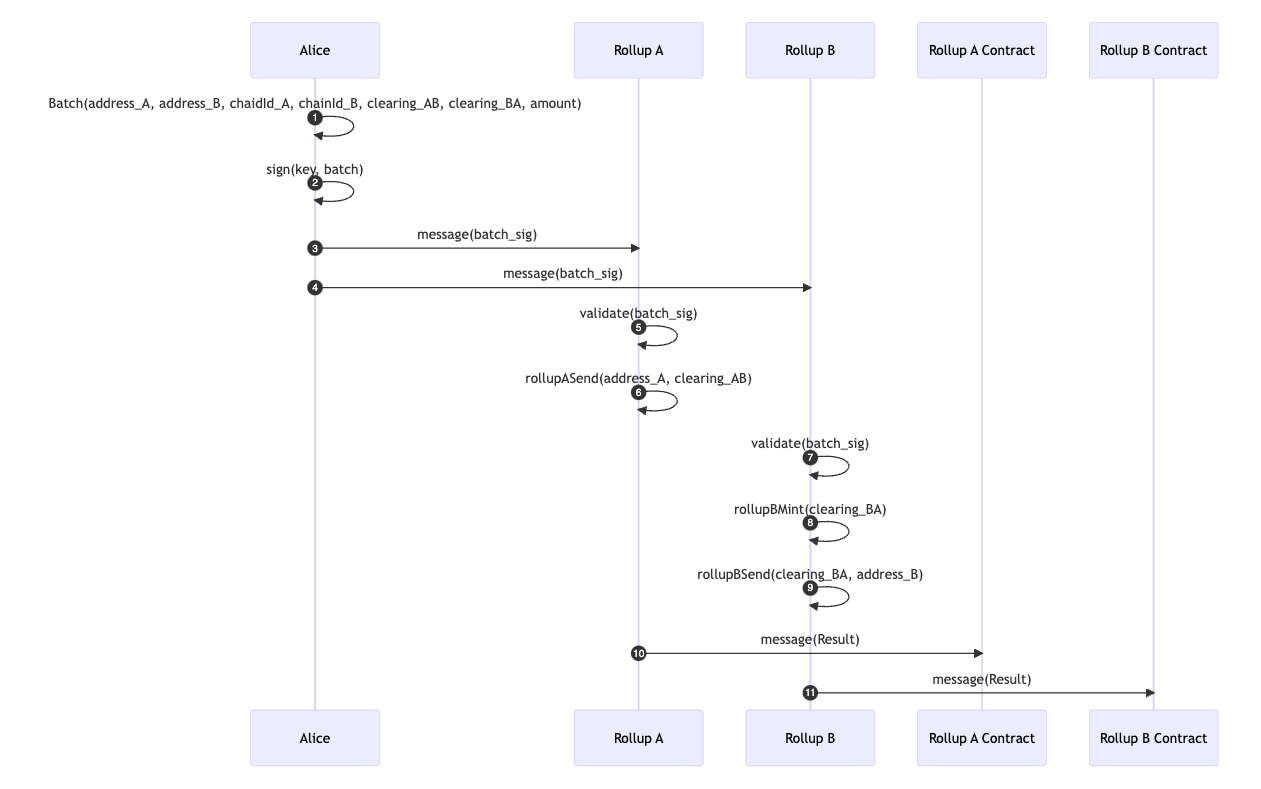}
\caption{Procedure for Inter-Rollup Transfers}
\label{fig_wide}
\end{figure}

Figure 5 depicts the procedure of transferring funds from a user of one rollup to a user of another rollup. As illustrated, two transaction types are necessary when using the inter-rollup transfer system. The first transaction type locks the sender's funds in the sending rollup, and the second type concerns the recipient rollup transferring funds to the receiver. The settlement rollup does not intervene during this process. Rollups record all transactions, and upon receiving a settlement request, they compile a list of relevant transactions to submit to the settlement rollup.

Initially, Alice, a user of Rollup A, employs the batch() function to create a bundle containing multiple transactions in order to transfer funds to Bob, a user of Rollup B. The bundle includes Alice's address (\(address_A\)), Bob's address (\(address_B\)), the sending rollup ID (\(chaidId_A\)), the receiving rollup ID (\(chainId_B\)), Rollup A's clearing account for B (\(clearing_{AB}\)), Rollup B's clearing account for A (\(clearing_{BA}\)), and the transfer amount (\(amount\)). The generated bundle comprises three transactions:

\begin{enumerate}
    \item rollupASend(): Alice transfers funds to Rollup A's clearing account for Rollup B
    \item rollupBMint(): Funds are issued from Rollup B's clearing account for Rollup A
    \item rollupBSend(): Funds are transferred from Rollup B's clearing account for Rollup A to Bob
\end{enumerate}

After creating the transactions, Alice uses her signing key (\(key\)) to sign the batch through the sign() function and forwards the signed transactions (\(batch_{sig}\)) to both rollups. Rollup A verifies Alice's status as a client and her possession of adequate funds for the transfer. Subsequently, Rollup A executes Alice's rollupASend() transaction.

Rollup B confirms Bob as a client and then executes the rollupBMint() and rollupBSend() transactions, issuing the specified amount of funds and transferring them to Bob's address.

Upon completing the transactions, both rollups modify their respective states. Following this, the rollups routinely compress the outcomes of the executed transactions and store them within the smart contracts of the rollups on the blockchain. It is assumed that the synchronization and finalization of the states for both rollups on the blockchain take place concurrently.

\subsection{Settlement for fund transfer}

\begin{figure}[H]
\centering
\includegraphics[width=1\linewidth]{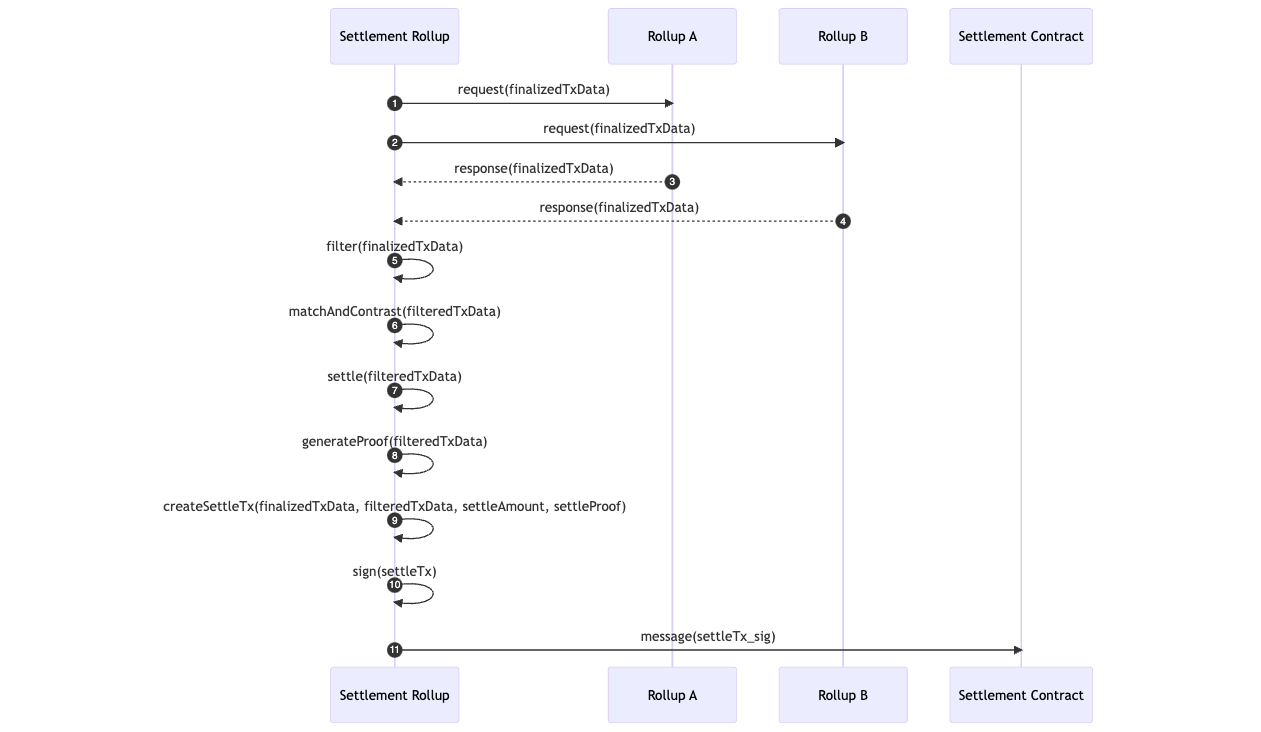}
\caption{Settlement of Inter-Rollup Transfer Transactions}
\label{fig_sim}
\end{figure}

Figure 6 depicts the settlement rollup's process of settling fund transfers between rollups. As demonstrated in the figure, the settlement rollup undertakes two distinct functions during this process. The first function consists of consolidating the comprehensive transaction history among rollups and computing the settlement balances for each respective bank. The second function involves transmitting the settlement results to the blockchain.

To initiate the settlement process, the settlement rollup first retrieves transaction information from the rollups. This information encompasses not only inter-rollup fund transfers but also all transaction data generated within the rollups. Moreover, the settlement rollup gathers data solely from transactions that have been recorded and finalized on the blockchain by each rollup.

Once the transaction information is collected, the settlement rollup filters specific transactions involving each rollup's clearing account using the filter() function. It then matches and contrasts the filtered transactions with matchAndContrast() function, ensuring that the sum of state changes for matched transactions remains zero. The settlement rollup subsequently calculates the settlement balance for each rollup using settle() function and generates a zero-knowledge proof to validate the settlement balance with generateProof() function (refer to section 4.2.3).

The settlement rollup employs createSettleTx() function to construct a transaction (\(settleTx\)) that enables the settlement smart contract to process the settlement balance. This transaction comprises the finalized transaction data (\(finalizedTxData\)) collected from the rollups, filtered transaction data (\(filteredTxData\)), the settlement balance (\(settleAmount\)), and a proof of validity for the settlement results (\(settleProof\)). Notably, the rollup transfer information data consists of a concatenated sequence of completed transaction data gathered from the rollups.

Lastly, the settlement rollup signs the transaction containing all these elements and submits the signed transaction (\(settleTx_{sig}\)) to the blockchain's settlement smart contract.

\subsection{Settlement in smart contract}

The settlement smart contract on the blockchain executes transactions to finalize the settlement of rollups. To accomplish this, the contract verifies the validity of the settlement results submitted by the settlement rollup and transfers funds via the rollups' settlement accounts. This procedure functions as described in section 4.3.

\section{Conclusion}

In this paper, we introduce an innovative solution to tackle the challenges of efficiency and usability in inter-rollup transfers within the Ethereum blockchain. By implementing a batch settlement method that consolidates multiple transactions for processing at once, our solution enhances the efficiency of transfers between rollups, as opposed to individually settling each transaction. Notably, we employ zero-knowledge proof algorithms to ensure the computational integrity of settlement rollups and guarantee system transparency and security by verifying balances and facilitating payments through Ethereum smart contracts.

Our proposed system addresses the scalability concerns of blockchain-based financial applications. By reducing the frequency of interactions with the blockchain during inter-rollup transfers, our approach increases throughput, processes settlements off-chain, and minimizes transfer latency for users. This method is anticipated to foster the effective utilization of blockchain-based financial services.

Nevertheless, there is room for improvement in the current research. In future studies, we intend to strengthen the safety, efficiency, and traceability of inter-rollup transfers by adopting standardized message formats for the seamless exchange of transfer transactions between users and rollups. Specifically, our goal is to securely deliver users' transfer transactions to rollups without exposing them to security risks, swiftly process intricate transactions between various rollups, and provide users and rollups with a clear understanding of transaction processes, making it easier to pinpoint the cause of issues.

\end{multicols}

\end{document}